\documentstyle[seceq,epsf,wrapft,twoside]{ptptex}
\newcommand{\rt}{\rightarrow}
\newcommand{\psip}{\psi(2S)}
\newcommand{\opi}{\omega\pi^+\pi^-}
\newcommand{\ok}{\omega K^+K^-}
\newcommand{\opp}{\omega p\bar p}  
\newcommand{\fipp}{\phi p\bar p}
\newcommand{\fikk}{\phi K^+ K^-}
\newcommand{\jpsi}{J/\psi}
\newcommand{\kk}{K^+K^-}
\newcommand{\fipi}{\phi\pi^+\pi^-}
\newcommand{\psipto}{\psi(2S)\rightarrow}

\newcommand{\EE}{e^+e^-}
\newcommand{\MM}{\mu^+\mu^-}
\newcommand{\PPJP}{\pi^+\pi^-J/\psi}

\newcommand{\pp}{\psi(2S)}

\newcommand{\etal}{\it et al.\rm}

\newcommand{\gpppr}{\gamma\pi^+\pi^-p\overline{p}}
\newcommand{\pppr}{\pi^+\pi^-p\overline{p}}
\newcommand{\chicJ}{\chi_{cJ}}
\newcommand{\gaa}{\gamma\Lambda\overline{\Lambda}}
\newcommand{\chicz}{\chi_{c0}}
\newcommand{\ppb}{p\overline{p}}
\newcommand{\chicJto}{\chi_{cJ} \rightarrow }
\newcommand{\aab}{\Lambda\overline{\Lambda}}
\newcommand{\chico}{\chi_{c1}}
\newcommand{\psp}{\psi(2S)}

\newcommand{\chictto}{\chi_{c2} \rightarrow }
\newcommand{\pspto}{\psi(2S) \rightarrow }
\newcommand{\ssb}{\Sigma^0\overline{\Sigma^0}}
\newcommand{\ra}{\rightarrow}
\newcommand{\psipp}{\pi^+\pi^- J/\psi}
\newcommand{\chict}{\chi_{c2}}

\newcommand{\BR}{{\cal B}}

\newcommand{\chiczto}{\chi_{c0} \rightarrow }

\notypesetlogo  
\title{Recent Charmonium Physics from BES
}
\author{%
Frederick A. {\sc Harris}$^1$ \\
for the BES Collaboration
}
\inst{%
$^{1}$ Department of Physics and Astronomy, 
The University of Hawaii\\
Honolulu, HI 96822
}
\recdate{%
May 16, 2003
}
\abst{%
Measurements of branching fractions for $\psip$ decays 
into $\opi$, $b_1\pi$, $\omega f_2(1270)$, $\ok$, $\opp$, $\fipi$, 
$\phi f_0(980)$, $\fikk$, and $\fipp$ final states, based
on a data sample of $(4.02\pm0.22)\times10^6 ~ \psip$
events collected with the BESI detector at the Beijing
Electron-Positron Collider, are reported. 
Using the same event sample,
radiative decays of the radially excited charmonium resonance, $\psi (2S),$
into $\pi \pi $, $K\overline{K}$ and $\eta \eta $ final states have been
measured.  The branching ratios
$B(\psi(2S)\rightarrow \gamma f_{2}(1270))=(2.12\pm 0.19\pm 0.32)\times
10^{-4}$ and
$B(\psi(2S)\rightarrow\gamma f_0(1710))\times B(f_0(1710)\rightarrow K^+K^-)
=(3.02\pm 0.45\pm 0.66)\times 10^{-5}$ are obtained.

Cross sections for $\EE \rightarrow$ $\EE$, hadrons, $\PPJP$, and
$\MM$ have been measured in the vicinity of the $\psip$ resonance
using the BESII detector.  The $\psip$ total width; partial widths to
hadrons, $\PPJP$, and muons; and corresponding branching fractions
have been determined to be $\Gamma_t=264 \pm 27 $ keV; $\Gamma_h=258
\pm 26 $ keV, $\Gamma_{\mu}=2.44 \pm 0.21$ keV, and
$\Gamma_{\PPJP}=85.4 \pm 8.7 $ keV; and $B_h=(97.79 \pm 0.15) \%$,
$B_{\PPJP}=(32.3\pm 1.4)\%$, $B_{\mu}=(0.93\pm 0.08) \%$,
respectively.

Decays of $J/\psi \rt \gamma \eta_c$ are used to determine
the mass and width of the $\eta_c$ using a sample of 58 M
$J/\psi$ events: $M_{\eta_c} = (2977.5 \pm 1.0 \pm 1.2)$ MeV and
$\Gamma_{\eta_c} = (17.0 \pm 3.7 \pm 7.4)$ MeV.

The first observation of $\chicJ$ (J=0,1,2) decays to $\aab$ is
  reported using $\psp$ data collected with the BESII detector at the
  BEPC. The branching ratios are determined to be \( \BR(\chicz
  \rightarrow \aab) = (4.7^{+1.3}_{-1.2} \pm 1.0)\times 10^{-4}\) , \(
  \BR(\chico \rightarrow \aab) = (2.6^{+1.0}_{-0.9} \pm 0.6)\times
  10^{-4}\) and \( \BR(\chict \rightarrow \aab) = (3.3^{+1.5}_{-1.3}
  \pm 0.7)\times 10^{-4}\).  Results are compared with model
  predictions.
}

\begin{document}
\maketitle

\setcounter{tocdepth}{4}

\section{Introduction}

The Beijing Spectrometer (BES) is a general purpose solenoidal
detector at the Beijing Electron Positron Collider (BEPC).  BEPC
operates in the center of mass energy range from 2 to 5 GeV with a
luminosity at the $J/\psi$ energy of approximately $ 5 \times 10^{30}$
cm$^{-2}$s$^{-1}$.  BES (BESI) is described in detail in Ref.
\citen{bes1}, and the upgraded BES detector (BESII) is described in
Ref. \citen{bes2}.  This paper presents some recent results; details
can be found in the references.

\section{\boldmath Hadronic $\psi(2S)$ decays}
\noindent
Both $J/\psi$ and $\psi(2S)$ decays to light hadrons are expected to 
proceed dominantly via $\psi \rt ggg $, with widths that are 
proportional to the square of the $c \overline{c}$ 
wave function at the origin.~\cite{appel} \  
This yields the expectation that
\begin{eqnarray}
Q_X = \frac{B(\psi(2S) \rt X_h)}{B(J/\psi \rt X_h)}  \approx  
\frac{B(\psi(2S) \rt e^+ e^-)}{B(J/\psi \rt e^+ e^-)} 
  \approx 12 \% \nonumber
\end{eqnarray}
It was first observed by MarkII \cite{mark2} that the
vector-pseudoscalar $\rho \pi$ and $K^*\overline{K}$ channels are
suppressed with respect to the $ 12 \% $ expectation - the ``$\rho
\pi$ puzzle''.  BES finds a $\rho
\pi$ suppression factor of $\sim$ 60; this and
many other BES $\psi(2S)$ branching ratio results can be found in
Refs.~\citen{rhopi,VT,eta,VP,bbbar}.

\begin{wraptable}{r}{7cm}
\begin{tabular}{lcc}
\hline
  Channel   & $B_{\psi(2S)\rightarrow X}$ & $Q_X (\%)$      \\ 
  ~~~~X     & ($10^{-4}$) & $(\%)$ \\ \hline
$\omega \pi^{+}\pi^{-}$   & $4.8\pm0.6\pm0.7$     
                          & $6.7\pm1.7$  \\ \hline 
$b_{1}^{\pm}\pi^{\mp}$    & $3.2\pm0.6\pm0.5$ 
                          & $11\pm3$ \\ \hline
$\omega f_{2}(1270)$      & $1.1\pm0.5\pm0.2$       
                          & $2.4\pm1.3$  \\
                          & $<1.5$                \\ \hline
$\omega K^{+}K^{-}$       & $1.5\pm0.3\pm0.2$           
                          & $20\pm8$   \\ \hline
$\omega p\bar{p}$         & $0.8\pm0.3\pm0.1$           
                          & $6.0\pm2.8$     \\ \hline 
$\phi\pi^+\pi^-$          & $1.5\pm0.2\pm0.2$             
                          & $18\pm5$    \\ \hline
$\phi f_0(980)$ \\
$\times (f_0\rightarrow\pi^+\pi^-)$ 
                          & $0.6\pm0.2\pm0.1$              \\
$\phi f_0(980)$      &  $1.1 \pm 0.4 \pm 0.1$
&$33\pm15$    \\ \hline
$\phi K^{+}K^{-}$         
                          & $0.6\pm0.2\pm0.1$        
                          & $7.3\pm2.6$      \\ \hline
$\phi p \bar{p}$          
                              & $<0.26$ &$<58$        \\ 
\hline
\end{tabular}\\[2pt]
\caption{ Branching fractions of $\psi(2S)$ and $Q_X$ values
for $\psi(2S)$ and
$J/\psi$ hadronic decays. The $B_{J/\psi}$ are taken from
the PDG. \cite{PDG} \  To determine $B(\phi f_0(980))$, we use
$B_{f_0\rightarrow\pi^+\pi^-}=0.521\pm0.016$ (PDG'96).}
\label{results}  
\end{wraptable}

\begin{figure}[htb]
  \parbox{\halftext}{
    
    \epsfxsize=6cm
     \centerline{\epsffile{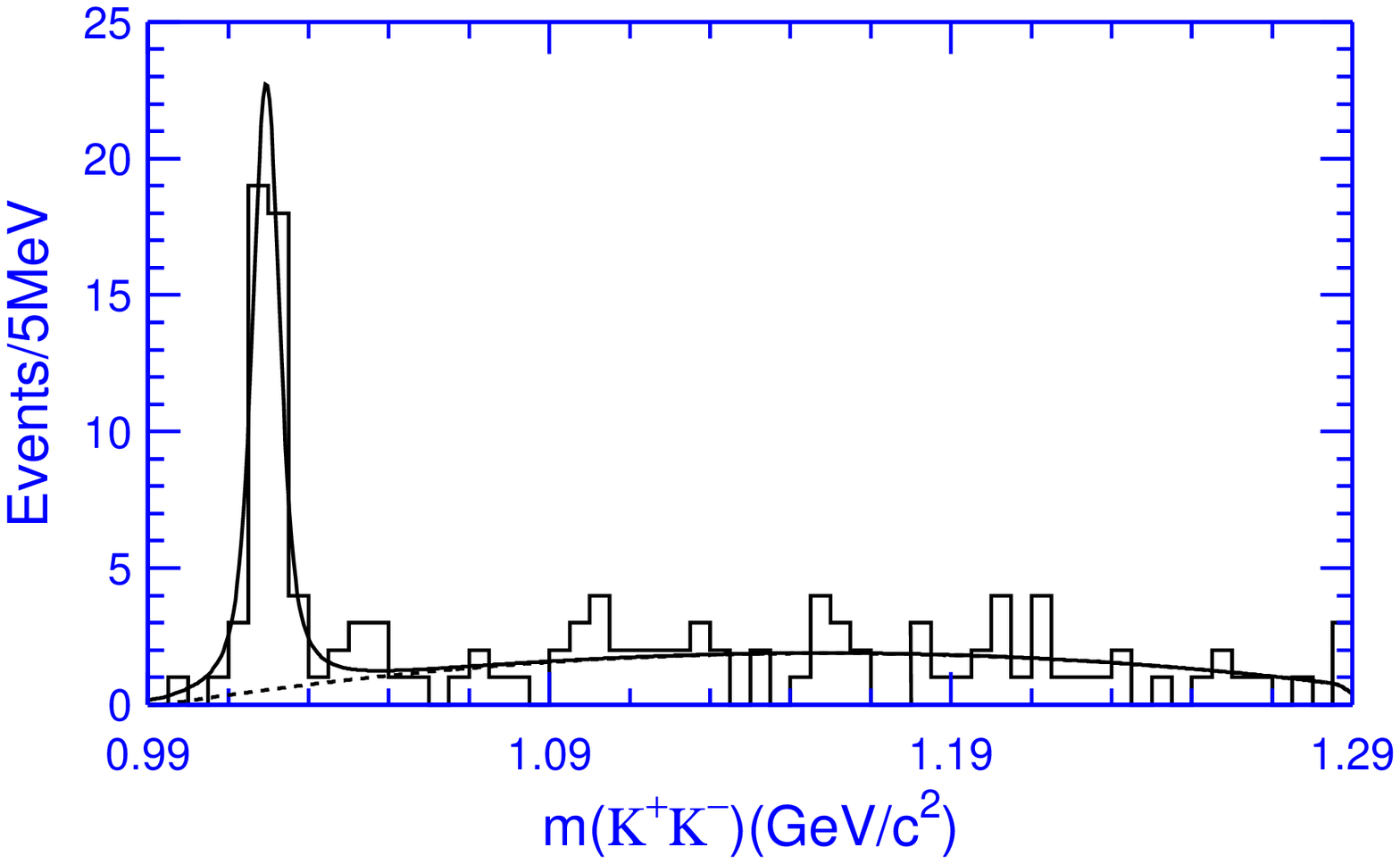}}
     \caption{The  $\kk$ invariant mass distribution for candidate
      $\psipto\fipi$ events. \label{fig:phipipifit} }}
   \hspace{8mm}
  \parbox{\halftext}{
    \epsfxsize=6cm
     \centerline{\epsffile{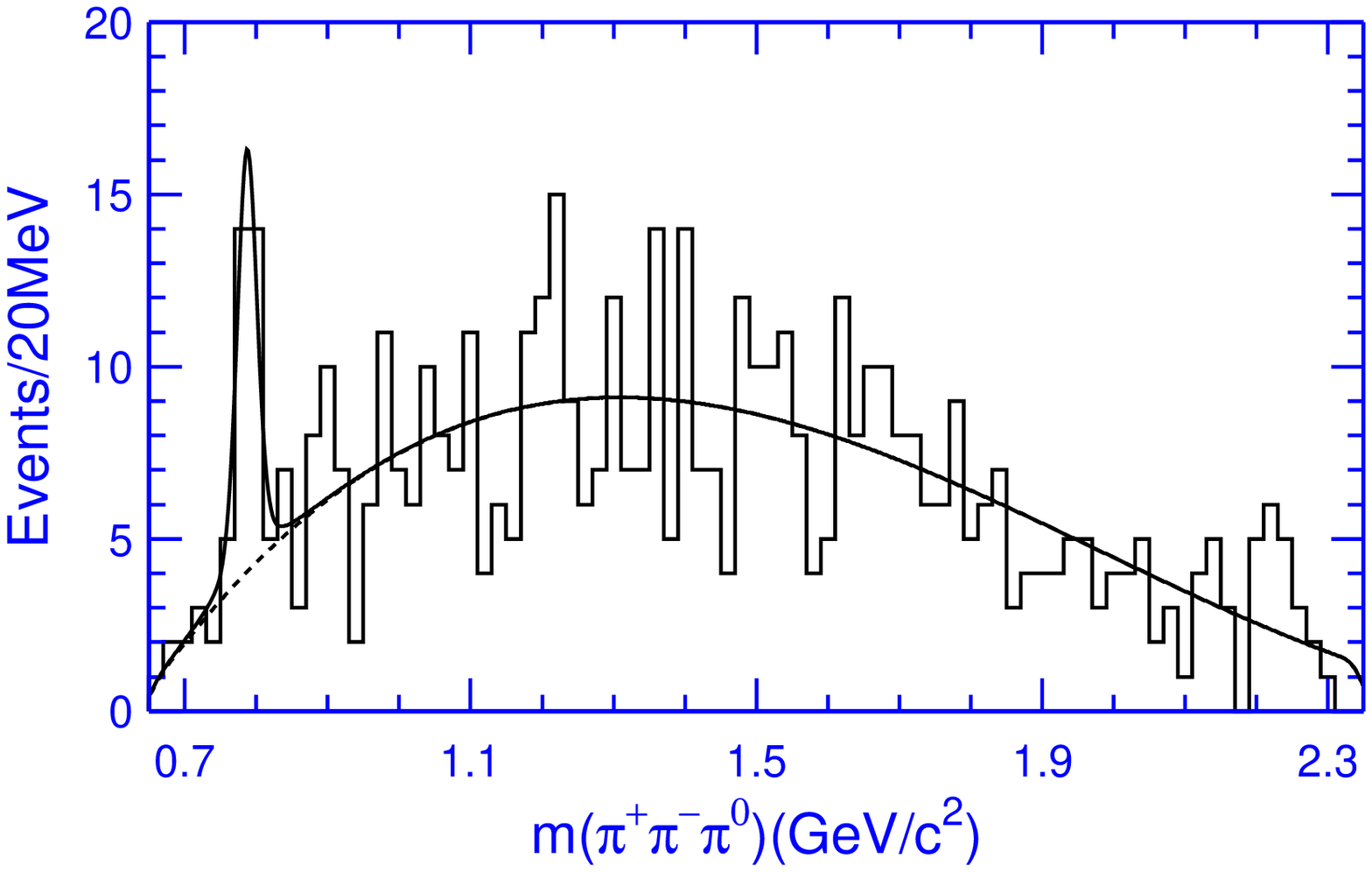}}
     \caption{The  $\pi^+\pi^-\pi^0$ mass distribution for candidate
      $\psipto\ok$ events. }
     \label{fig:wkkfit}}
\end{figure}

Here, we report measurements of branching fractions for $\psip$ decays
involving an $\omega$ or a $\phi$, including $\opi$, $b_1\pi$, $\omega
f_2(1270)$, $\ok$, $\opp$, $\fipi$, $\phi f_0(980)$, $\fikk$, and
$\fipp$ final states, based on a data sample of
$(4.02\pm0.22)\times10^6 ~ \psip$ events collected with the BESI
detector at the Beijing Electron-Positron Collider. Events are
selected using particle identification and kinematic fitting.
As an example, the
$\kk$ invariant mass distribution for candidate $\psipto\fipi$ events
is shown in Fig.~\ref{fig:phipipifit}, where a clear $\phi$ peak can
be seen. In Fig.~\ref{fig:wkkfit}, the $\pi^+ \pi^- \pi^0$ mass
distribution for $\psip\rt\ok$ events is shown;  there is a clear
$\omega$ peak.   
We obtain the branching ratios and $Q_X$ values
shown in Table~\ref{results}.
The branching fractions for $b_1\pi$ and  
$\omega f_2(1270)$ update previous BES results, while those for other 
decay modes are first measurements. The ratios of $\psip$ and $\jpsi$ 
branching fractions are smaller than the expected 12\% rule
by a factor of six for $\omega f_2(1270)$, 
by a factor of two for $\opi$, $\opp$, and $\fikk$,  
while for other studied channels the ratios are 
consistent with expectations within errors. 
For more detail on this analysis, see Ref.~\citen{la1}.


In perturbative QCD, the radiative $J/\psi $ and $\psi (2S)$ decays should
be similar to hadronic decays except instead of decaying into three gluons,
the radiative mode decays via two gluons and one photon.
Thus one power of the coefficient $\alpha _{S}$ is replaced by
$\alpha _{QED}$ in the cross section formula, and it is expected that the
``12\%'' rule should also work for radiative decay modes \cite{Radiative12rule}.
Hence the ratio of $B(\psi(2S)\rightarrow \gamma X)$ to
$B(J/\psi \rightarrow \gamma X)$ for different final states $X$ should be
roughly 12\%.

\begin{figure}[htb]
  \parbox{\halftext}{
    \epsfxsize=6cm
     \centerline{\epsffile{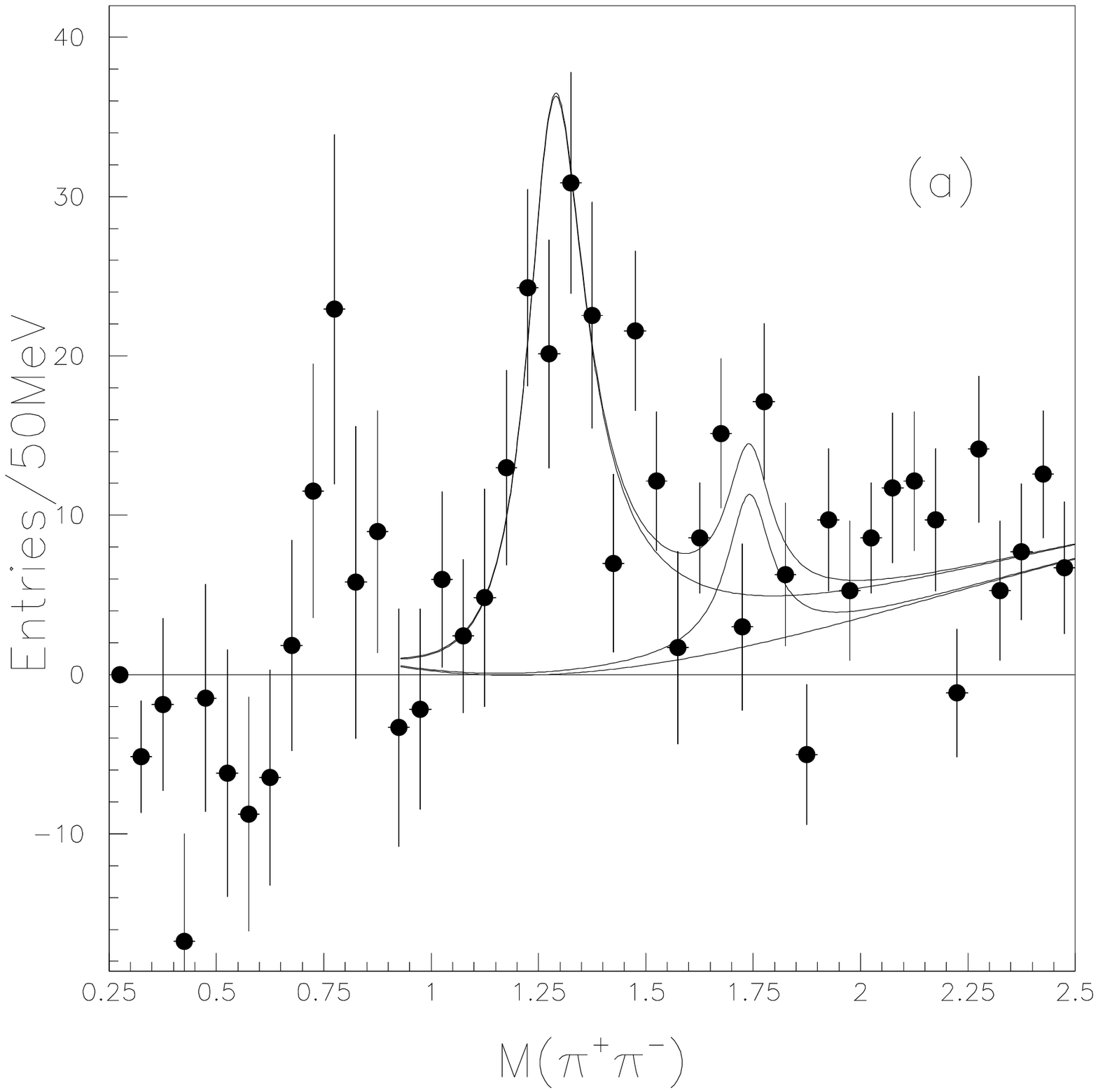}}}
   \hspace{8mm}
  \parbox{\halftext}{
    \epsfxsize=6cm
     \centerline{\epsffile{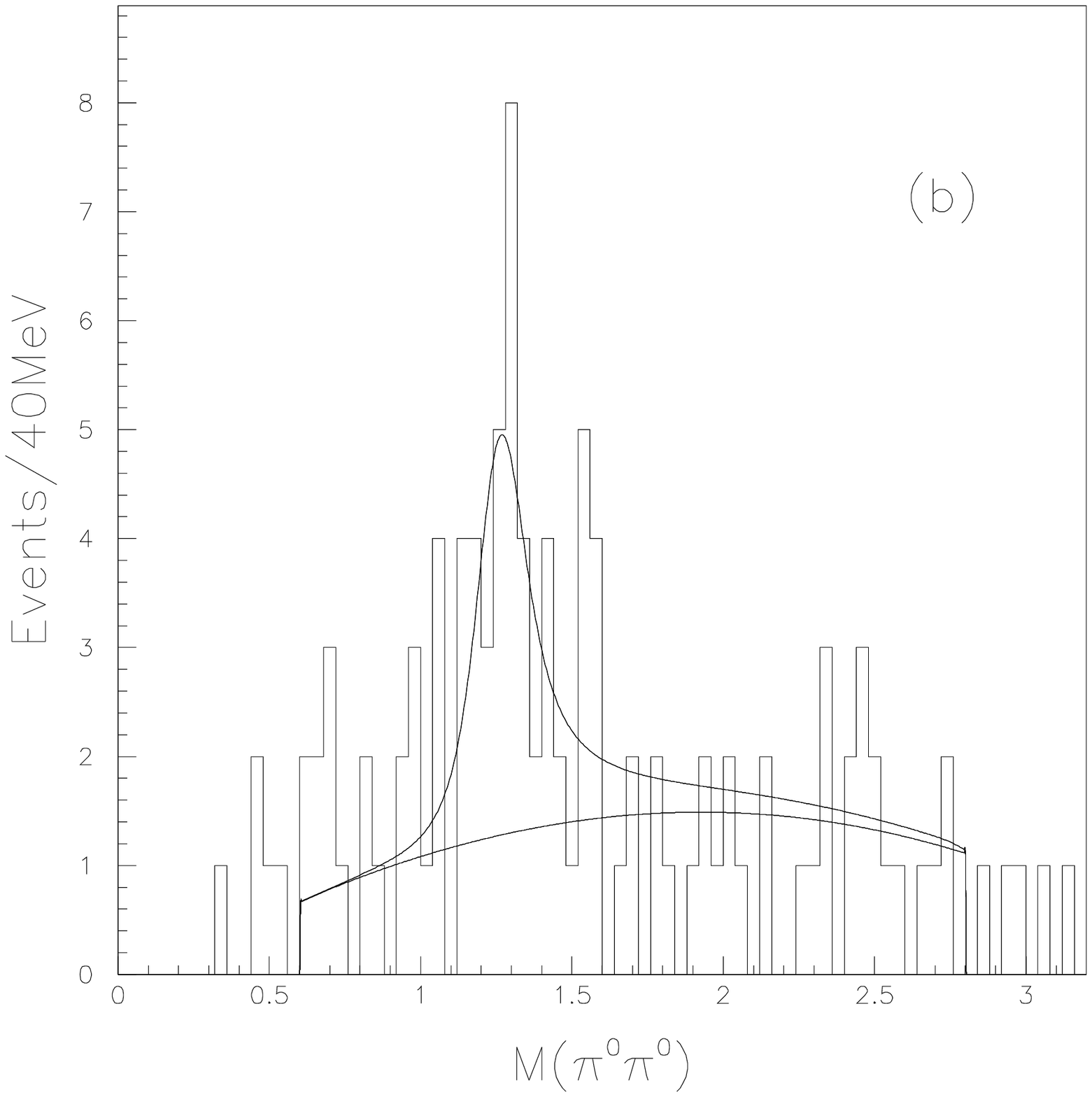}}}
 \caption{\label{fig:pipi}
          (a): $M_{\pi^+\pi^-}$ fit result.
               The four curves presented in 
               the figure are the following: a background curve, a 
               Breit-Wigner function to describe the $f_2(1270)$ on top of the 
               background, a Breit-Wigner function to describe the $f_0(1710)$ 
               on top of the background, and the total of the two 
               Breit-Wigners and the background. The fitting range is 0.9 GeV 
               to 2.5 GeV, since there is some $\rho$ background
               below 0.9 GeV. The background at higher mass is due to 
               processes such as $\psi(2S)
               \rightarrow$ neutrals $J/\psi$, 
               $J/\psi\rightarrow \pi^+\pi^-\pi^0 $ .
          (b): $M_{\pi^0\pi^0}$ fit result. The curves shown are a 
               Breit-Wigner to describe the $f_2(1270)$ and a polynomial to 
               describe the background.
         }

\end{figure}

Here we report measurements of branching fractions for $\psi(2S)
\rightarrow \gamma \pi^+\pi^-$, $\gamma \pi^0 \pi^0$, $\gamma K^+K^-$,
$\gamma K^0_S K^0_S$, and $\gamma \eta\eta$. The  $\pi \pi$ invariant
mass distributions for  $\psi(2S)
\rightarrow \gamma \pi \pi$ are shown in Fig.~\ref{fig:pipi}, where
a clear $ f_{2}(1270)$ is seen.  Results are summarized in
Tables~\ref{tab:12rule} through \ref{tab:bf_ratios_22}.
First measurements of the $\psi(2S)\rightarrow \gamma f_{2}(1270)$ and
$\psi (2S)\rightarrow \gamma f_0(1710)\rightarrow \gamma K^+K^-$ and
$\gamma K_S^0 K_S^0$ branching fractions are given. 
A clear $f_0(1710)$ signal
in $\psi(2S)$ radiative decay into $K^+K^-$ final states is observed.
The results are consistent with the ``12\%'' rule.
In addition, first measurements of the
branching fractions of $\chi_{c0}$ and $\chi_{c2}$ decay into $\pi^0\pi^0$,
$\chi_{c0}$ decay into $\eta\eta$, and an upper limit of the branching
fraction of $\chi_{c2}$ decay into $\eta\eta $ are reported (see Table
\ref{tab:bf_ratios_22}).  For more detail, see Ref.~\citen{yangwei}.

\begin{table}[htb] \centering
 \begin{tabular}{|l|l|l|}   \hline
 Final state & 
               $B(\psi(2S)\rightarrow)(\times 10^{-4})$ &
               $B(\psi(2S))/B(J/\psi)$ \\ \hline\hline
 $\gamma f_2(1270)$ &
               $2.12\pm 0.19\pm 0.32$ &
               $(15.4\pm 3.1)$\% \\ \hline
 $\gamma f_0(1710)\rightarrow\gamma K^+K^-$ &
               $0.302\pm 0.045\pm 0.066$ &
               $(7.1^{+2.1}_{-2.0})$\% \\ \hline
 \end{tabular}
 \caption{Values for $B(\psi(2S)\rightarrow) \gamma f_2(1270)$ and
 $B(\psi(2S)\rightarrow)  K^+K^-$ and comparison with the 12\% rule.}
 \label{tab:12rule}
\end{table}

\begin{table}[htb] \centering
 \begin{tabular}{|l|l|} \hline
  Mode & $B(\times 10^{-4})$  \\ \hline\hline
  $\psi(2S)\rightarrow\gamma f_2(1270)$ from $\gamma\pi^+\pi^-$ &
            $2.08\pm 0.19\pm 0.33$  \\ \hline
  $\psi(2S)\rightarrow\gamma f_2(1270)$ from $\gamma\pi^0\pi^0$ &
            $2.90\pm 1.08\pm 1.07$  \\ \hline
  $\psi(2S)\rightarrow\gamma f_2(1270)$ from $\gamma\pi\pi$ &
            $2.12\pm 0.19\pm 0.32$  \\ \hline
  $\psi(2S)\rightarrow\gamma f_0(1710)\rightarrow\gamma\pi\pi$ from 
                                              $\gamma\pi^+\pi^-$ &
            $0.301\pm 0.041\pm 0.124$  \\ \hline
  $\psi(2S)\rightarrow\gamma f_0(1710)\rightarrow\gamma K^+K^-$ &
            $0.302\pm 0.045\pm 0.066$  \\ \hline
  $\psi(2S)\rightarrow\gamma f_0(1710)\rightarrow\gamma K^0_SK^0_S$ &
            $0.206\pm 0.094\pm 0.108$  \\ \hline
 \end{tabular}
 \caption{Branching fractions for
          $\psi(2S)\rightarrow\gamma X\rightarrow\gamma P\overline{P}$
          modes ($P$ stands for pseudo-scalar). }
 \label{tab:bf_ratios_12}
\end{table}

\begin{table}[htb] \centering
 \begin{tabular}{|l|l|l|}   \hline
  Mode & $B(\times 10^{-3})$ & 
  $B\times B(\psi(2S)\rightarrow\gamma\chi_{c0,2})$
  \\
                  &   &  $ (\times 10^{-4})$ \\ \hline\hline
  $\chi_{c0}\rightarrow\pi^0\pi^0$ & 
    $2.79\pm 0.32\pm 0.57$ & $2.42\pm 0.28\pm 0.44$ \\ \hline
  $\chi_{c2}\rightarrow\pi^0\pi^0$ &
    $0.98\pm 0.27\pm 0.56$ & $0.67\pm 0.19\pm 0.38$ \\ \hline
  $\chi_{c0}\rightarrow\eta\eta$ & 
    $2.02\pm 0.84\pm 0.59$ & $1.76\pm 0.73\pm 0.49$ \\ \hline
  $\chi_{c2}\rightarrow\eta\eta$ & 
    $<1.37$                & $<0.93$          \\ \hline
 \end{tabular}
 \caption{The $\chi_c$ decay branching fractions
          for $\chi_{c0,2}\rightarrow\pi^0\pi^0$ or $\eta\eta$. }
 \label{tab:bf_ratios_22}
\end{table}

\newpage
\section{\boldmath BES $\psi(2S)$ Scan Results}

\begin{wrapfigure}{r}{6.7cm}
\epsfxsize=6.0cm
\centerline{\epsffile{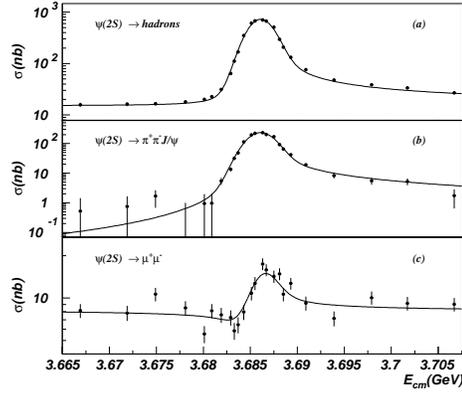}}
\caption{The cross section for (a) $ e^+ e^- \rightarrow$ hadrons, (b)
$ e^+ e^- \rightarrow$ $\pi^+\pi^- J/\psi$, and (c)
$ e^+ e^- \rightarrow \mu^+ \mu^- $ versus center-of-mass energy. 
The solid curves represent the results of the fit to the data.}
\label{fig4c}
\end{wrapfigure}

In 1999, after the $R$-scan, \cite{rscan} \ BES did a careful
$\psi(2S)$ scan.  The purpose was to improve the accuracies of the
$\psi(2S)$ parameters: the total width ($\Gamma_t$), and partial widths
into hadrons ($\Gamma_h$), $\mu^+ \mu^-$ ($\Gamma_{\mu}$), and $\pi^+
\pi^- J/\psi$ ($\Gamma_{\pi^+ \pi^- J/\psi}$), and the corresponding
branching fractions, $B(h)$, $B(\mu)$, and $B(\pi^+ \pi^-
J/\psi)$. $B(\pi^+ \pi^- J/\psi)$ and $B(\mu)$ are important because
these decays are used to identify $\pp$ in B decays ($B \rt \pp
K^0_S$).

A total of 24 energy points between 3.67 and 3.71 GeV were scanned.
The total integrated luminosity was 760 nb$^{-1}$.  We assume
$\Gamma_t = \Gamma_h + \Gamma_{\mu} + \Gamma_e + \Gamma_{\tau}$, along
with lepton universality: $\Gamma_e = \Gamma_{\mu} =
\Gamma_{\tau}/0.38847$.  The cross sections versus scan point energy
and fit curves are shown in Fig.~\ref{fig4c}, and the fit
results are given in Table~\ref{fitres}.  We obtain a first
measurement of $\Gamma_{\pi^+ \pi^- J/\psi}$, and $B(h)$, $B(\mu)$,
and $B(\pi^+ \pi^- J/\psi)$ have improved precision compared to the
PDG values. \cite{PDG} \  The value for $\Gamma_t$ agrees within
errors with a
previous BES value of ($252 \pm 37$) keV. \cite{tautau}
A complete description of this work can be found in Ref.~\citen{la2}.

\begin{wraptable}{l}{7cm}
\doublerulesep 0.5pt
 \begin{tabular}{c||c|c} \hline \hline
      Value     &   BES        &    PDG2002             \\ \hline \hline
$\Gamma_t$(keV) &$264\pm 27$   &$300 \pm 25$     \\ 
$\Gamma_h$(keV) &$258\pm 26$   &                 \\ 
$\Gamma_{\pi\pi J/\psi}$(keV)
                &$85.4 \pm 8.7$&                 \\
$\Gamma_{\mu}$(keV)
                &$2.44\pm0.21$ &                 \\
${\cal B}_{h}(\%)$
                &$97.79\pm0.15$&$98.10\pm0.30$   \\ 
${\cal B}_{\pi\pi J/\psi}(\%)$
                &$32.3\pm1.4 $ &$30.5\pm 1.6$    \\ 
${\cal B}_{\mu}(\%)$
                &$0.93\pm0.08$ &$0.7\pm0.09$     \\ \hline \hline
\end{tabular} \\ 
\caption{\label{fitres} $\psi(2S)$ scan results and comparison with the PDG2002.
 \cite{PDG} \
$\Gamma_{\mu}$ value given using the  assumption $\Gamma_e=\Gamma_{\mu}$.}
\end{wraptable}

\newpage
\section{\boldmath   $\eta_c$ Parameters}

The mass and width of the $\eta_c$ are rather poorly known; the
confidence level for the PDG weighted average mass is only 0.001.
\cite{PDG} \ Previously BES measured the $\eta_c$ mass using the BESI
4.02 M $\psi(2S)$ sample and obtained $M_{\eta_c} = (2975.8 \pm 3.9
\pm 1.2)~{\rm MeV}$. \cite{etac1} BES also used 7.8 M BESI $J/\psi$
events and obtained $M_{\eta_c} = (2976.6 \pm 2.9 \pm 1.3)$ MeV.
\cite{etac2} \  For the two data sets combined, $M_{\eta_c} = (2976.3
\pm 2.3 \pm 1.2)$ MeV and the total width $\Gamma_{\eta_c} = (11.0 \pm
8.1 \pm 4.1)$ MeV. {\cite{etac2}

Here, the mass and width have been determined using our BESII 58 M $J/\psi$
event sample.  We use the channels $J/\psi \rt \gamma \eta_c$, with $\eta_c
\rt p \bar{p}$, $K^+ K^- \pi^+ \pi^-$, $\pi^+ \pi^- \pi^+ \pi^-$,
$K^{\pm} K^o_S \pi^{\mp}$, and $\phi \phi$.  Events are selected using
particle identification and kinematic fitting. Figs.~\ref{fig:eta1} and
\ref{fig:eta2} show the mass distributions in the $\eta_c$ mass region for
$J/\psi \rt \gamma \eta_c$, $\eta_c \rt p \bar{p}$ and $\eta_c \rt
K^+ K^- \pi^+ \pi^-$, respectively.  Combining the five decay
channels, we obtain  $M_{\eta_c} = (2977.5
\pm 1.0 \pm 1.2)$ MeV and $\Gamma_{\eta_c} = (17.0 \pm 3.7 \pm 7.4)$
MeV, to be compared to the current PDG values: $M_{\eta_c} = (2979.7
\pm 1.5)$ MeV and $\Gamma_{\eta_c} = (16.0^{+3.6}_{-3.2})$ MeV.
\cite{PDG} \  The results for the mass and width are compared with
previous measurements, including previous BES measurements, in
Figs.~\ref{fig:mass} and \ref{fig:width}.  The results are in good
agreement with previous BES measurements and the PDG fit values.
More detail on this analysis can be found in Ref. \citen{etanew}.

\begin{figure}[htb]
  \parbox{\halftext}{
    \epsfxsize=6cm
     \centerline{\epsffile{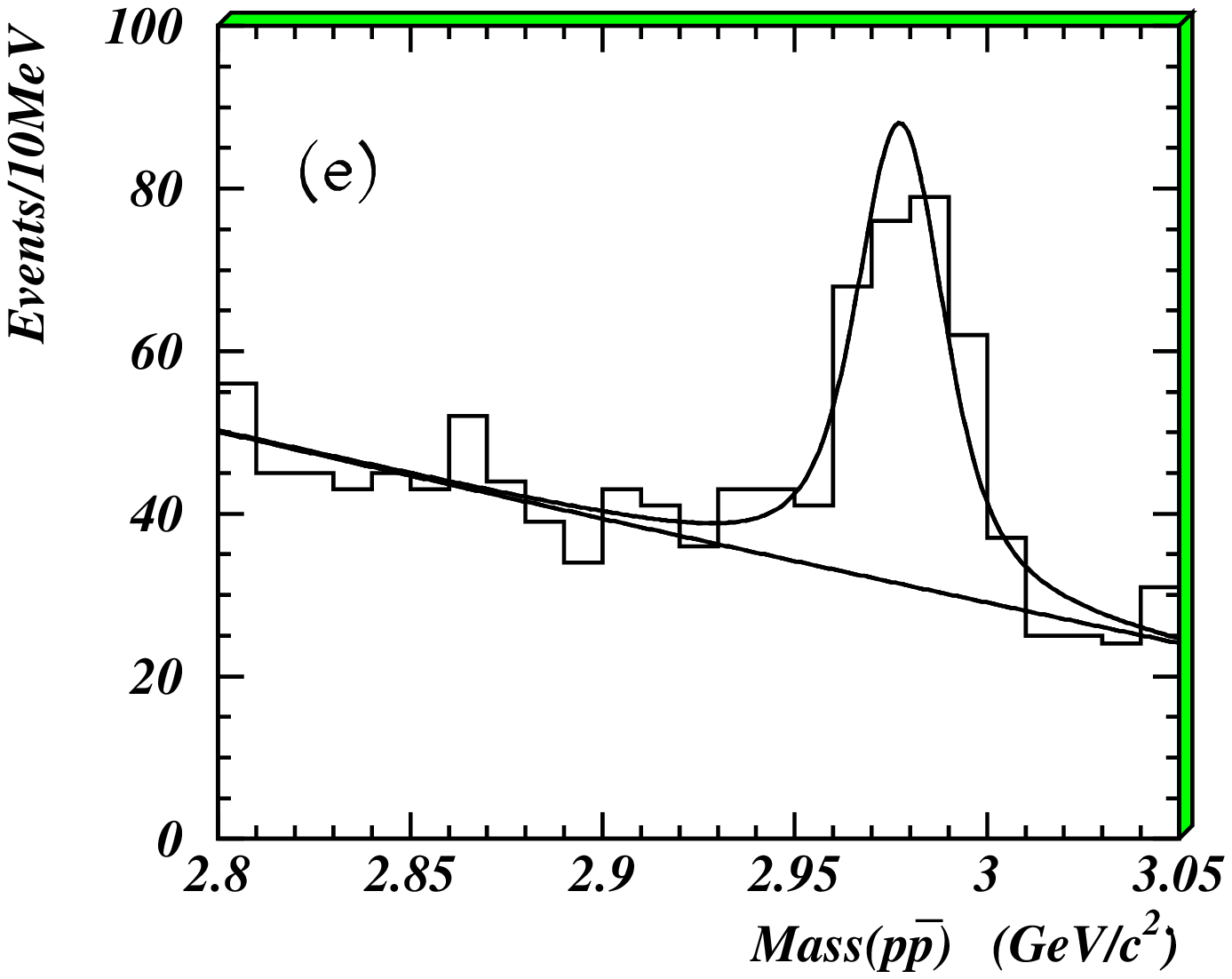}}
     \caption{The  $m_{p\bar{p}}$ invariant mass distribution in the
       $\eta_c$ region.}
     \label{fig:eta1}}
   \hspace{8mm}
  \parbox{\halftext}{
    \epsfxsize=6cm
     \centerline{\epsffile{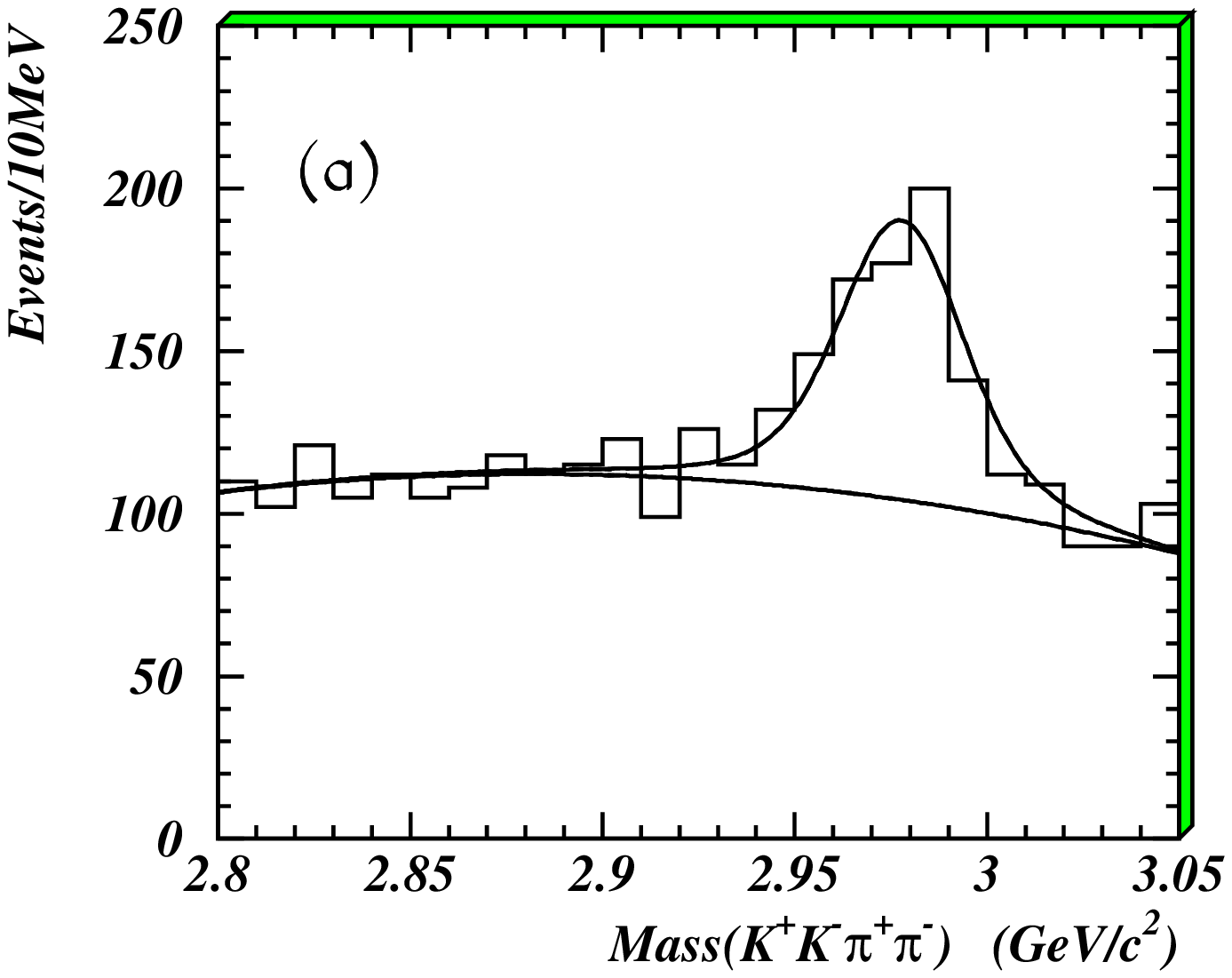}}
     \caption{The  $m_{K^+K^-\pi^+\pi^-}$ invariant mass distribution
       in the $\eta_c$ region.}
     \label{fig:eta2}}
\end{figure}

\begin{figure}[htb]
  \parbox{\halftext}{
    \epsfxsize=6cm
     \centerline{\epsffile{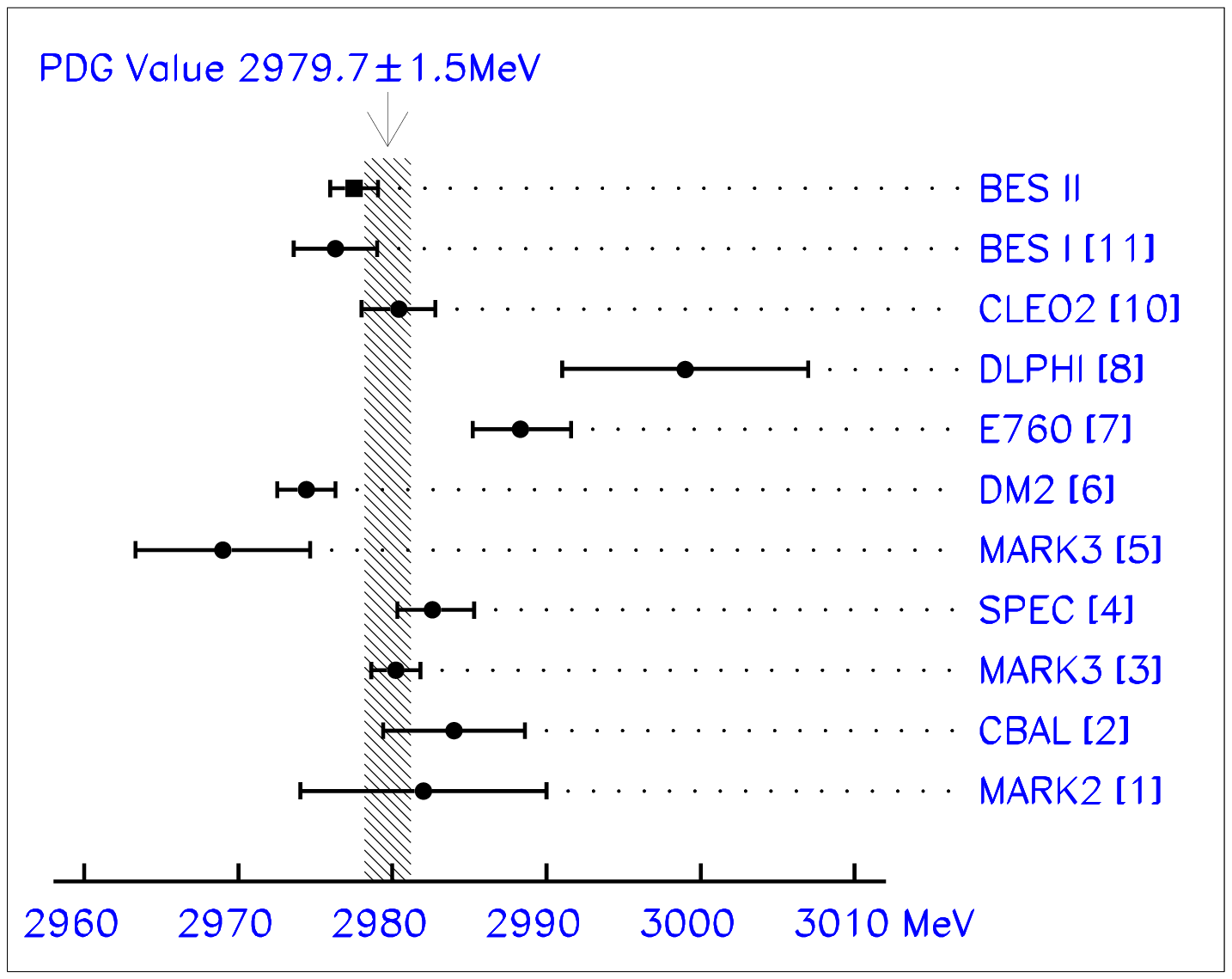}}
     \caption{Mass measurements of the $\eta_c$ meson.}
     \label{fig:mass}}
   \hspace{8mm}
  \parbox{\halftext}{
    \epsfxsize=6cm
     \centerline{\epsffile{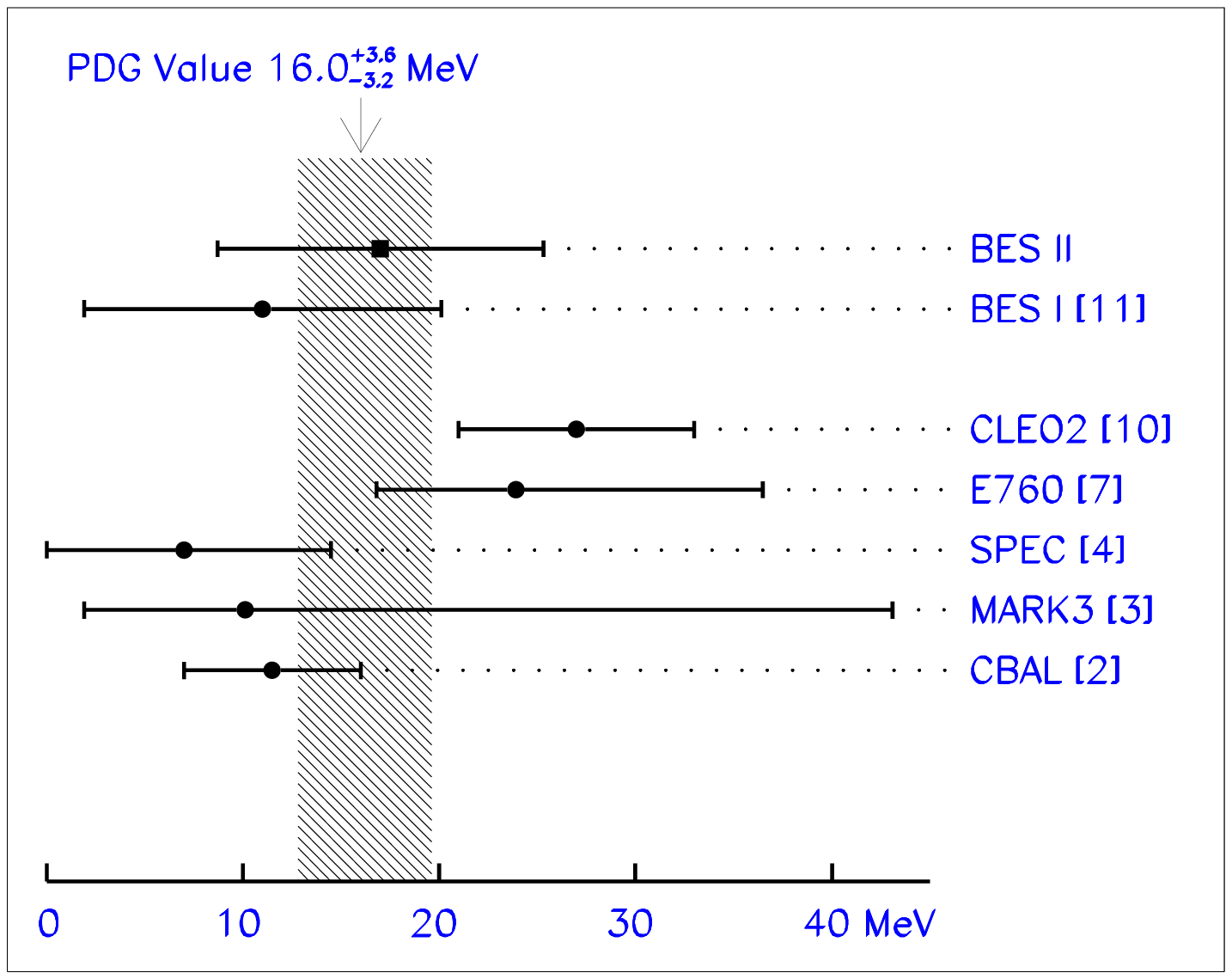}}
     \caption{Width measurements of the $\eta_c$ meson.}
     \label{fig:width}}
\end{figure}

\section{\boldmath $\chi_J \rt \Lambda \overline{\Lambda}$}

It has been shown both in theoretical calculations and experimental
measurements that the lowest Fock state expansion (color singlet
mechanism, CSM) of charmonium states is insufficient to describe
P-wave quarkonium decays. Instead, the next higher Fock state (color
octet mechanism, COM) plays an important role.~\cite{so,width} \  Our
earlier measurement~\cite{width} of the total width of the $\chicz$ agrees
rather well with the COM expectation.  The calculation of the partial
width of $\chicJto \ppb$, by taking into account the COM of $\chicJ$
decays and using a carefully constructed nucleon wave
function, ~\cite{wong} \ obtains results in reasonable agreement with
measurements. ~\cite{PDG} \  The nucleon wave function was then
generalized to other baryons, and the partial widths of many other
baryon anti-baryon pairs predicted. Among these predictions, the
partial width of $\chicJto \aab$ is about half of that of $\chicJto
\ppb$~(J=1,2).~\cite{wong}

\begin{figure}[htb]
  \parbox{\halftext}{
    \epsfxsize=6cm
     \centerline{\epsffile{ma-ma.epsi}}
\caption{Scatter plot of $\pi^+\overline{p}$ versus $\pi^-p$ invariant
mass for selected $\gpppr$ events with the $\pppr$ mass in the $\chicJ$ mass 
region.}
\label{ma-ma}}
   \hspace{8mm}
  \parbox{\halftext}{
    \epsfxsize=6cm
     \centerline{\epsffile{ma.epsi}}
\caption{Mass distribution of $\pi^+\overline{p}$ ($\pi^-p$)
recoiling against a $\Lambda$ ($\overline{\Lambda}$)
(mass $<1.15$~GeV) for events in the $\chicJ$ mass region.
Dots with error bars are data and the histogram is the
Monte Carlo simulation, normalized to the $\Lambda$ signal
region (two entries per event).}
\label{ma}}
\end{figure}

Fig.~\ref{ma-ma} shows a scatter plot of the $\pi^+\overline{p}$
versus the $\pi^-p$ invariant mass for events with $\pppr$ mass
between $3.38$~GeV/$c^2$ and $3.60$~GeV/$c^2$, using the BESII 15
million $\psi(2S)$ event sample. The cluster of events
in the lower left corner shows a clear $\aab$ signal.
Selecting events in $\chicJ$ mass region and requiring the mass
of $\pi^+\overline{p}$ ($\pi^-p$) to be smaller than 1.15~GeV/$c^2$, the
$\pi^-p$ ($\pi^+\overline{p}$) mass distribution shown in
Fig.~\ref{ma} is obtained. A clear $\Lambda$ signal can be seen, and the 
background below the peak is very small.

\begin{wrapfigure}{r}{6.5cm}
\epsfxsize=6.5cm
\centerline{\epsffile{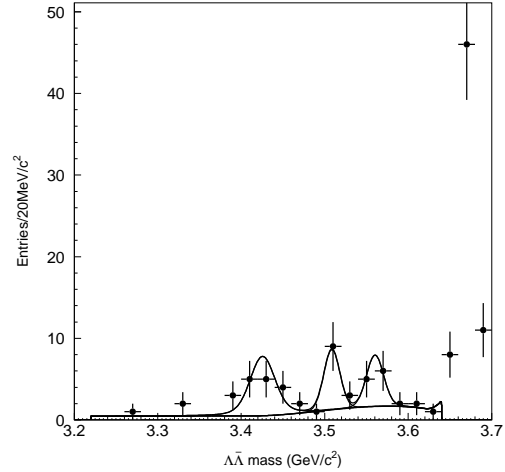}}
\caption{Mass distribution of $\gaa$ candidates fitted with three
resolution smeared Breit-Wigner functions and background, as
described in the text.}
\label{maafit}
\end{wrapfigure}

After requiring that both the $\pi^+\overline{p}$ and the $\pi^-p$
mass lie within twice the mass resolution around the nominal $\Lambda$
mass, the $\aab$ invariant mass distribution shown in
Fig.~\ref{maafit} is obtained. There are clear $\chicz$, $\chico$, and
$\chictto \aab$ signals. The highest peak around the $\psp$ mass is
due to $\pspto \aab$ with a fake photon.

Background from non $\aab$ events is estimated from the $\Lambda$ mass
sidebands, and this can be described in fitting the $\aab$ mass
spectrum by a linear background.  The background from channels with
$\aab$ production, including $\pspto \aab$, $\pspto \ssb$, $\pspto
\Lambda \overline{\Sigma^0} + c.c.$, $\pspto \Xi^0 \overline{\Sigma^0}
+ c.c.$, $\pspto \gamma \chicJ, \chicJto \Sigma^0\overline{\Sigma^0}
\ra \gamma \gamma \aab$, and $\pspto \psipp \rightarrow \pppr$, are
simulated by Monte Carlo.

Fixing the $\chicz$, $\chico$ and $\chict$ mass resolutions at their
Monte Carlo predicted values, and fixing the widths of the three
$\chicJ$ states to their world average values, ~\cite{PDG} \ the mass
spectrum (Fig.~\ref{maafit}) was fit with three Breit-Wigner functions
folded with Gaussian resolutions and background, including a linear
term representing the non $\aab$ background and a component
representing the $\aab$ background. The unbinned maximum likelihood
method was used to fit the events with $\aab$ mass between 3.22 and
3.64~$\hbox{GeV}/c^2$, and a likelihood probability of 27\% was
obtained, indicating a reliable fit.  Fig.~\ref{maafit} shows the fit
result, and the fitted masses are $(3425.6\pm 6.3)\hbox{MeV}/c^2$,
$(3508.5\pm 3.9)~\hbox{MeV}/c^2$ and $(3560.3\pm 4.6)~\hbox{MeV}/c^2$
for $\chicz$, $\chico$ and $\chict$, respectively, in agreement with
the world average values. ~\cite{PDG}
The branching ratios of $\chicJto \aab$ obtained are
\[ \BR(\chicz \rightarrow \aab)
     = (4.7^{+1.3}_{-1.2} \pm 1.0)\times 10^{-4} ,\]
\[ \BR(\chico \rightarrow \aab)
     = (2.6^{+1.0}_{-0.9} \pm 0.6)\times 10^{-4} ,\]
\[ \BR(\chict \rightarrow \aab)
     = (3.3^{+1.5}_{-1.3} \pm 0.7)\times 10^{-4} ,\]
where the first errors are statistical and the second are 
systematic. 

Compared with the corresponding branching ratios of $\chicJto
\ppb$, ~\cite{PDG} \ the branching ratios of 
$\chico$ and $\chictto \aab$ agree with the corresponding 
branching ratios to $\ppb$ within two sigma. This is somewhat 
in contradiction with the expectations from Ref.~\citen{wong},
although the errors are large.

As for $\chiczto \aab$, the measured value agrees with the $\ppb$
measurements from BES and E835~\cite{width,e835} within 2 standard
deviations. One should also note that there is no prediction for
$\BR(\chiczto \aab)$.   More detail may be
found in Ref. \citen{gll}.

\section{Summary}
Branching fractions are determined, many for the first time, using the
4.2 million BESI $\psi(2S)$ event sample.  They are used to test the
``12 \%'' rule.  Results from a fit to a careful scan in the vicinity
of the $\psi(2S)$ are presented.  The 58 million BESII $J/\psi$ event
sample is used to measure the mass and width of the $\eta_c$.
Finally,
$\aab$ events are observed for the first time in $\chicJ$ decays using
the BESII 15 million $\psp$ event sample, and corresponding branching ratios 
are determined. 

\acknowledgements

The author would like to thank Prof. S. Ishida, Prof.
K. Takamatsu, and all the other members of the Sigma Group for their
support during the Symposium.

\end{document}